\documentclass[preprint,aps,showpacs,preprintnumbers,amsmath,amssymb,superscriptaddress]{revtex4}

\usepackage{graphicx}
\usepackage{dcolumn}
\usepackage{bm}

\begin{document}


\title{A BCS wavefunction approach to the BEC-BCS crossover of exciton-polariton condensates}

\author{Tim Byrnes}
\affiliation{National Institute of Informatics, 2-1-2
Hitotsubashi, Chiyoda-ku, Tokyo 101-8430, Japan}
\affiliation{Institute of Industrial Science,
University of Tokyo, 4-6-1 Komaba, Meguro-ku, Tokyo 153-8505,
Japan} 

\author{Tomoyuki Horikiri}
\affiliation{E. L. Ginzton Laboratory, Stanford University, Stanford, CA 94305}

\author{Natsuko Ishida}
\affiliation{National Institute of Informatics, 2-1-2
Hitotsubashi, Chiyoda-ku, Tokyo 101-8430, Japan}
\affiliation{Department of Information and Communication Engineering, The University of Tokyo, 7-3-1 Hongo, Bunkyo-ku, Tokyo 113-8656, Japan} 

\author{Yoshihisa Yamamoto}
\affiliation{National Institute of Informatics, 2-1-2
Hitotsubashi, Chiyoda-ku, Tokyo 101-8430, Japan}
\affiliation{E. L. Ginzton Laboratory, Stanford University, Stanford, CA 94305}

\date{\today}

\date{\today}

\begin{abstract}
The crossover between low and high density regimes of exciton-polariton condensates is examined using a BCS wavefunction approach.
Our approach is an extension of the BEC-BCS crossover theory for excitons, but includes a cavity
photon field. The approach can describe both the low density limit, where the system can 
be described as a Bose-Einstein condensate (BEC) of exciton-polaritons, and the high density limit, where
the system enters a photon dominated regime. In contrast to the exciton BEC-BCS crossover where the system approaches an electron-hole plasma, 
the polariton high density limit has strongly correlated electron-hole pairs. At intermediate densities, there is a regime with BCS-like properties, with a peak at non-zero momentum of the singlet pair function. 
We calculate the expected photoluminescence and give several experimental signatures of the crossover.
\end{abstract}

\maketitle

In recent years there is an increasing consensus that a Bose-Einstein condensate (BEC) of exciton-polaritons has been realized experimentally \cite{kasprzak06,deng02}.
For exciton BECs, it is well known that the system crosses over into a BCS state of weakly correlated electrons and holes at high density  \cite{kjeldysh68,comte82}. 
A natural question is then: Does a BEC-BCS crossover also occur for exciton-polaritons? Littlewood and co-workers have examined this question and have predicted that with increasing
density the system transitions from a BEC state to a photon BEC state \cite{eastham01,keeling04,keeling05,keeling07}. In an intermediate density regime under suitable conditions they predict a BCS-like regime \cite{keeling05}. 
The model that they deal with is a model of non-interacting excitons coupled to a 
common photonic cavity. In this model, the excitons do not contain an internal electron hole structure and are treated as 
spins localized on lattice sites. Our purpose here is to include the electron and hole components
as well as their Coulomb interaction.  By doing so we find that several new effects are present which have implications
on experimentally observable quantities. 

\begin{figure}
\scalebox{0.3}{\includegraphics[bb=0 0 1420 886]{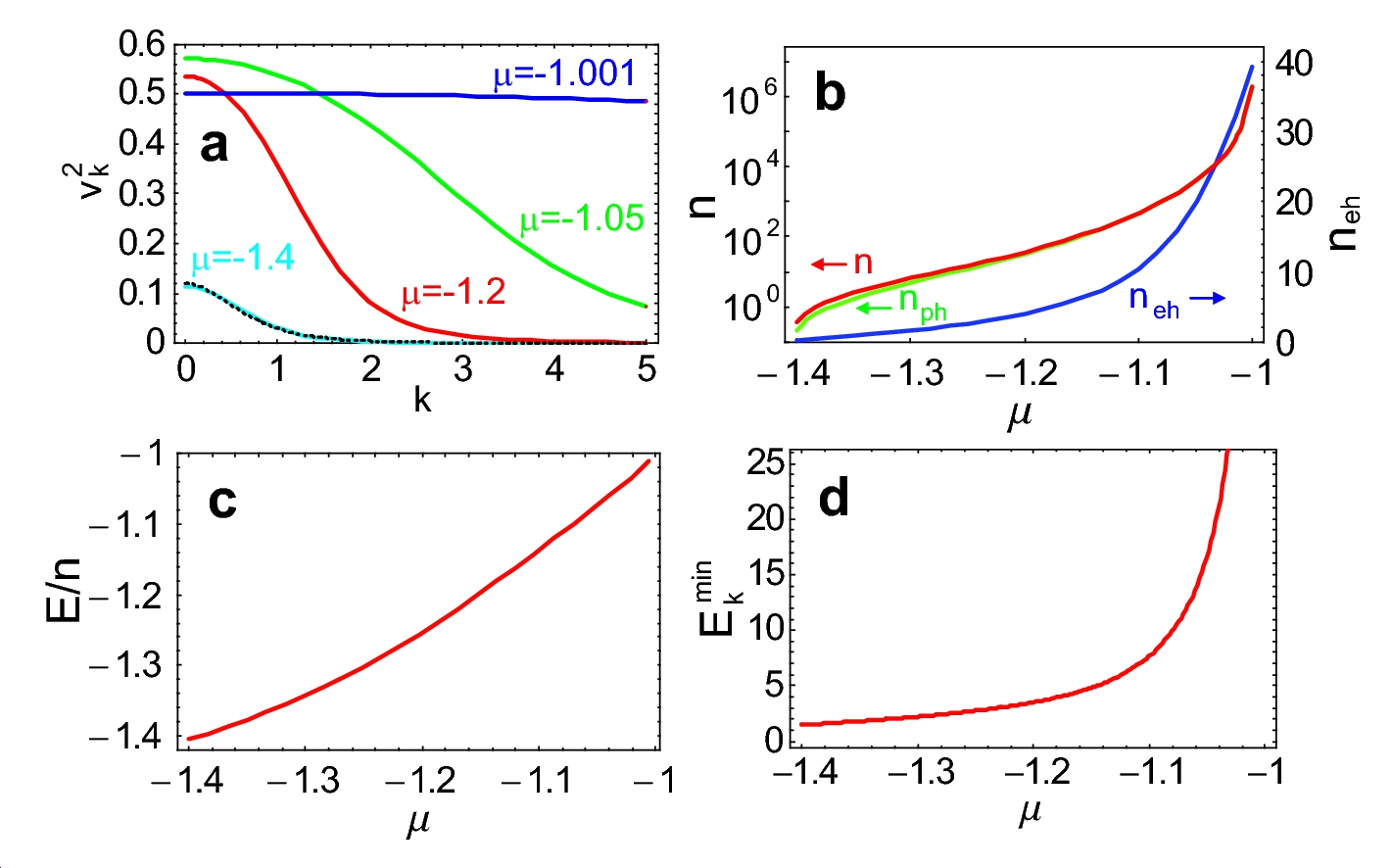}}
\caption{\label{fig1}
(a) The occupation probability $ v_{\bm{k}}^2 $ of electron hole pairs of momentum $ k $ for a cavity photon energy of $ \omega = -1 $ and photon coupling $ \Omega = 0.1 $.  Dotted line shows the 1s exciton wavefunction for comparison.  The (b) electron-hole pair, photon, and total particle density, (c) energy per electron-hole pair, (d) gap energy versus chemical potential. }
\end{figure}

The Hamiltonian we consider is
\begin{align}
H & = H_{\mbox{\tiny kin}}+ H_{\mbox{\tiny Coul}} + H_{\mbox{\tiny ph}} - \mu n  \label{grandham} \\
H_{\mbox{\tiny kin}} & = \sum_{\bm{k}} \left[ \frac{\hbar^2 k^2}{2 m_e} e_{\bm{k}}^\dagger e_{\bm{k}} +  \frac{\hbar^2 k^2}{2 m_h}h_{\bm{k}}^\dagger h_{\bm{k}} \right]  \nonumber \\
H_{\mbox{\tiny Coul}} & =  \frac{1}{2} \sum_{\bm{k},\bm{k}',\bm{q}} V(\bm{q}) \Big[ e_{\bm{k}+\bm{q}}^\dagger e_{\bm{k}'-\bm{q}}^\dagger e_{\bm{k}'} e_{\bm{k}}
+ h_{\bm{k}+\bm{q}}^\dagger h_{\bm{k}'-\bm{q}}^\dagger h_{\bm{k}'} h_{\bm{k}} \nonumber \\
& - 2 e_{\bm{k}+\bm{q}}^\dagger h_{\bm{k}'-\bm{q}}^\dagger h_{\bm{k}'} e_{\bm{k}} \Big] \nonumber \\
H_{\mbox{\tiny ph}} & = \Omega  \sum_{\bm{k}} \left[ e_{\bm{k}}^\dagger h_{-\bm{k}}^\dagger a + a^\dagger h_{-\bm{k}} e_{\bm{k}} \right] +  \omega a^\dagger a \nonumber ,
\end{align}
where $ e_{\bm{k}} $ and $ h_{\bm{k}} $ are the fermion annihilation operators for electrons and holes with momentum $ \bm{k} $, $ V(\bm{q}) = \frac{e^2}{2 \epsilon q} $ is the Coulomb interaction, $m_e $ and $ m_h $ are the electron and hole effective masses, $ a $ is the cavity photon annihilation operator, $ \omega $ is the cavity photon energy, $ \Omega $ is the coupling strength of the cavity photon to the electrons and holes and $ n = a^\dagger a + \sum_{\bm{k}} \left[ e_{\bm{k}}^\dagger e_{\bm{k}} + h_{\bm{k}}^\dagger h_{\bm{k}} \right] $ 
is the total particle number operator. We introduce a chemical potential $ \mu $ into the Hamiltonian in order to fix the total particle number. 

We use a BCS wavefunction ansatz of the form
\begin{equation}
\label{bcswavefunction}
|\Phi \rangle = \exp[\lambda a^\dagger-\lambda^2/2] \prod_k \left[ u_{\bm{k}} + v_{\bm{k}} e_{\bm{k}}^\dagger h_{-\bm{k}}^\dagger \right] | 0 \rangle ,
\end{equation}
where $ u_{\bm{k}}^2 + v_{\bm{k}}^2 = 1 $ in analogy to the BCS wavefunction used for excitons \cite{kjeldysh68,comte82}.  An additional 
coherent state photonic term is included of the same form as in Ref. \cite{eastham01}. The photon density is $ n_{\mbox{\tiny ph}} = \lambda^2 $ and the electron-hole density is $ n_{\mbox{\tiny eh}} = \sum_{\bm{k}} v_{\bm{k}}^2 $, giving a total particle density $ n = n_{\mbox{\tiny ph}} + n_{\mbox{\tiny eh}} $.

The BCS wavefunction (\ref{bcswavefunction}) is equivalent to the solution of the Hamiltonian (\ref{grandham}) by mean field theory \cite{tinkham96}.  
Following the mean field derivation, we assume a pairing Hamiltonian in eq. (\ref{grandham}) and restrict $ \bm{k}' = - \bm{k} $ in the attractive Coulomb interaction.
In the repulsive Coulomb terms we assume a Hartree Fock approximation, and restrict $ \bm{q} = \bm{k}' -\bm{k} $. Hartree terms (corresponding to $ \bm{q} =0 $), are not present due to the assumption of charge neutrality \cite{comte82}. Expanding the operators $ e_{\bm{k}}^\dagger e_{\bm{k}} $, $ h_{\bm{k}}^\dagger h_{\bm{k}} $, $ e_{\bm{k}}^\dagger h_{-\bm{k}}^\dagger $, and $ a $ around their mean values and keeping only linear terms gives
\begin{align}
H & = E_0 +  \Omega D ( a + a^\dagger)  + (\omega -\mu) a^\dagger a + \sum_{\bm{k}} \xi_{\bm{k}} ( e_{\bm{k}}^\dagger e_{\bm{k}} +  h_{\bm{k}}^\dagger h_{\bm{k}} ) \nonumber \\
& - \sum_{\bm{k}} ({\Delta_{\bm{k}}} - \Omega \lambda ) 
( e_{\bm{k}}^\dagger h_{-\bm{k}}^\dagger + h_{-\bm{k}} e_{\bm{k}}) 
\end{align}
where
\begin{align}
\xi_{\bm{k}} & = \frac{\hbar^2 k^2}{2 m} -\mu - X_{\bm{k}} \nonumber \\
X_{\bm{k}} & = \sum_{\bm{k}'}  V(\bm{\bm{k}-\bm{k}'}) \langle e_{\bm{k}'}^\dagger e_{\bm{k}'} \rangle \\
\Delta_{\bm{k}} & = \sum_{\bm{k}'}  V(\bm{\bm{k}-\bm{k}'}) \langle e_{\bm{k}'}^\dagger h_{-\bm{k}'}^\dagger \rangle  \\
D & = \sum_{\bm{k}} \langle e_{\bm{k}}^\dagger h_{-\bm{k}}^\dagger \rangle \\
E_0 & = \sum_{\bm{k}} \left[ X_{\bm{k}} \langle e_{\bm{k}}^\dagger e_{\bm{k}} \rangle
+ \Delta_{\bm{k}} \langle e_{\bm{k}}^\dagger h_{-\bm{k}}^\dagger \rangle \right] - 2\Omega D \lambda
\end{align}
and we have taken all expectation values as real, $ 2/m = 1/m_e + 1/m_h $, $ \lambda = \langle a \rangle $, and $ \langle e_{\bm{k}}^\dagger e_{\bm{k}} \rangle = \langle h_{-\bm{k}}^\dagger h_{-\bm{k}} \rangle $ due to charge neutrality. The photonic part may be diagonalized by introducing the operator $ B^\dagger = a^\dagger - c $, 
and demanding that the off-diagonal terms vanish. This gives the condition
\begin{equation}
\label{lambdaeqn}
\lambda= - \frac{\Omega D}{\omega -\mu } .
\end{equation}
The remaining part of the Hamiltonian may be diagonalized by a transformation
\begin{align}
e_{\bm{k}} & = u_{\bm{k}} \gamma_{\bm{k}0} + v_{\bm{k}} \gamma_{\bm{k}1}^\dagger \\
h_{\bm{k}}^\dagger & = - v_{\bm{k}} \gamma_{\bm{k}0} + u_{\bm{k}} \gamma_{\bm{k}1}^\dagger .
\end{align}
Demanding that the off-diagonal terms disappear, we obtain the Hamiltonian
\begin{equation}
\label{finalham}
H = \epsilon_0 + (\omega -\mu) B^\dagger B + \sum_{\bm{k}} E_{\bm{k}} \left( \gamma_{\bm{k}0}^\dagger  \gamma_{\bm{k}0} + \gamma_{\bm{k}1}^\dagger  \gamma_{\bm{k}1} \right), 
\end{equation}
where
\begin{align}
\epsilon_0 & = \sum_{\bm{k}} \left[  \xi_{\bm{k}} - E_{\bm{k}}
+ \Delta_{\bm{k}} \langle e_{\bm{k}}^\dagger h_{-\bm{k}}^\dagger \rangle + X_{\bm{k}} \langle e_{\bm{k}}^\dagger e_{\bm{k}}  \rangle \right] + \frac{\Omega^2 D^2}{\omega-\mu} .
\nonumber
\end{align}
The parameters $ u_{\bm{k}} $ and $ v_{\bm{k}} $ satisfy the standard BCS algebra \cite{tinkham96}  
\begin{align}
\label{bcs1}
\langle e_{\bm{k}}^\dagger e_{\bm{k}} \rangle = v_{\bm{k}}^2 = \frac{1}{2}( 1 - \frac{\xi_{\bm{k}}}{E_{\bm{k}}}) \\
\label{bcs2}
\langle e_{\bm{k}}^\dagger h_{-\bm{k}}^\dagger \rangle =  u_{\bm{k}}  v_{\bm{k}} =  \frac{\Delta_{\bm{k}} - \Omega \lambda }{2 E_{\bm{k}}} \\
E_{\bm{k}} = \sqrt{\xi_{\bm{k}}^2 + ({\Delta_{\bm{k}}} - \Omega \lambda )^2} .
\label{bcsequations}
\end{align}
The gap energy $ E_{\bm{k}}^{\mbox{\tiny min}} $ is defined to be the value of  $ E_{\bm{k}} $ minimized over
all momenta $ k $. In our numerical results, we use units such that the momentum is measured in units of $ 1/a_B = me^2/4 \pi \epsilon \hbar^2 $, the energy is 
measured in units $ E_0 = e^2/4 \pi \epsilon a_B $ (SI units used throughout).  In the low density limit with no photon field, the exciton energy is $ E/E_0 = -1 $.  Zero detuning therefore corresponds to a photon energy of $ \omega/E_0 = -1 $ (the photon energy is negative since we measure energies relative to the bandgap energy). The equations
(\ref{bcs1})-(\ref{bcsequations}) with (\ref{lambdaeqn}) are solved self-consistently to obtain our results.

In Fig. \ref{fig1}a the occupation probability $ v_{\bm{k}}^2 $ of electron
hole pairs is shown.  In the low density limit ($\mu \approx -1.4 $), the distribution coincides with the exciton wavefunction $ v_{\bm{k}} \propto 1/(1+k^2)^2$ \cite{comte82}. Furthermore, the density of electron-hole pairs and photons is nearly equal in this limit (Fig. \ref{fig1}b).  Fig. \ref{fig1}c
shows the energy per electron-hole pair, which approaches an energy of $ E/E_0 \approx -1.4 $. The lowering of the energy is due to the strong coupling and anti-crossing of the exciton and photon to form
a lower polariton.  
Finally, the gap energy shown in Fig. \ref{fig1}d is equal to the energy of a polariton $ E_{\bm{k}}^{\mbox{\tiny min}} \approx 1.4$. The gap energy in this case is the energy required to turn a polariton into a free electron hole pair. We thus conclude that exciton-polaritons are correctly reproduced in the low density limit.

As the density is increased, Fig. \ref{fig1}a reveals that the momentum distribution spreads out to higher momentum states. This is precisely the opposite behavior to what is expected in the excitonic BEC-BCS crossover. In a standard BCS state, the instability towards forming a Cooper pair is weakened with increasing density, because the
electron-hole attraction becomes increasingly screened by the surrounding electrons and holes. This results in a $ v_{\bm{k}} $ distribution that approaches a Fermi step for high density, for the excitonic case.  In the case of exciton-polaritons, the electron-hole attraction in fact becomes {\it enhanced} with increasing density.  This 
is evidenced by the gap energy which increases with density in Fig. \ref{fig1}d, instead of decreasing in the exciton case. 

What is the origin for this enhanced attraction? Fig. \ref{fig1}b reveals that at high density the photon number 
greatly exceeds that of the electron-hole number. This can be explained due to a difference in the particle statistics
of the two excitations. Photons are true bosonic particles, hence any number of them can be excited with an energy 
cost $ \omega $. Meanwhile, electron-hole pairs are fermions, and suffers a phase space filling effect. In order to excite more fermionic particles, electrons and holes of increasingly higher momenta and energy must be occupied in order to increase the particle number. Thus it is favorable to excite photons rather than electron hole pairs to minimize the total energy, explaining the large imbalance in these numbers.

\begin{figure}
\scalebox{0.3}{\includegraphics[bb=0 0 1342 892]{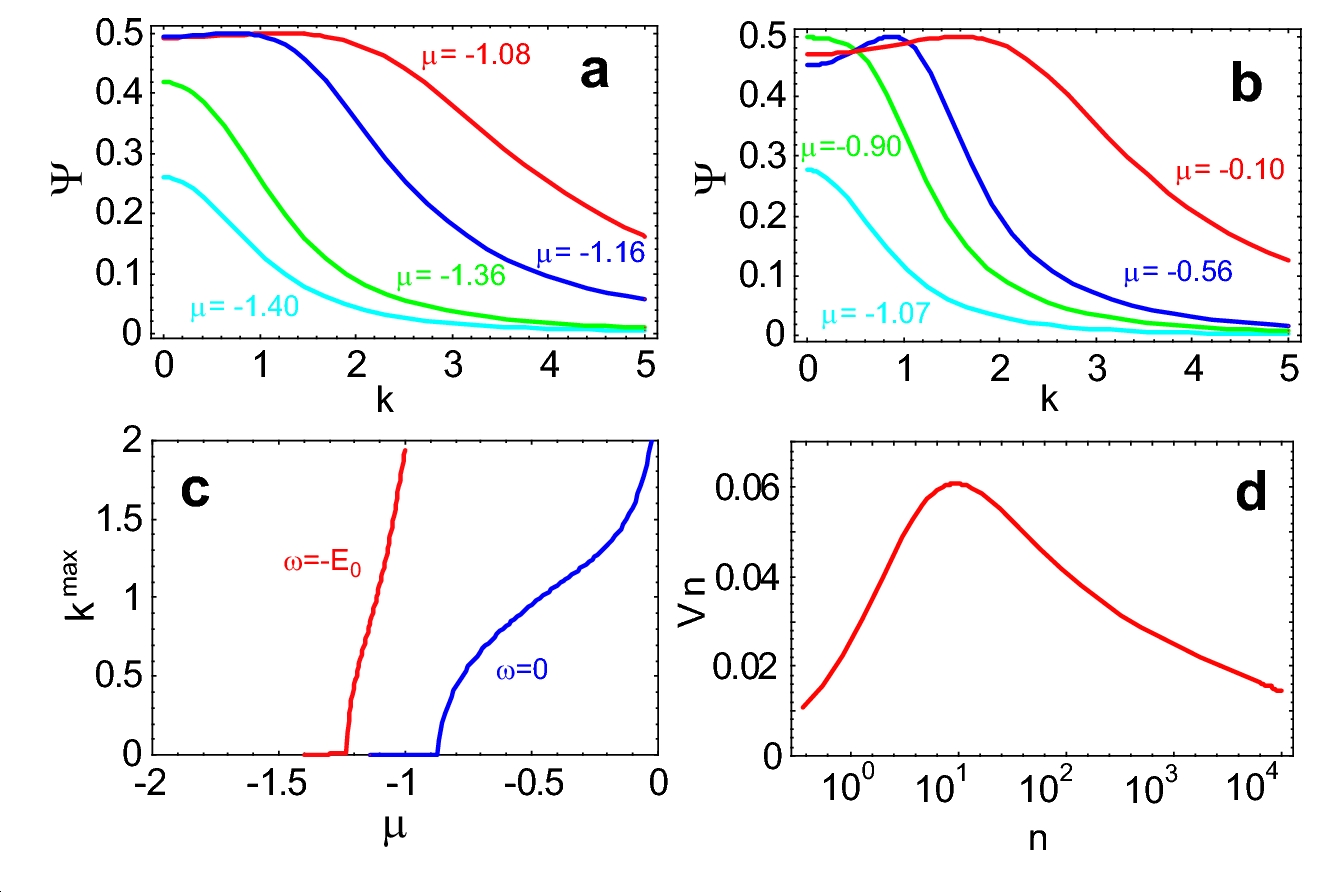}}
\caption{\label{fig2}
The singlet pair function $ \Psi(\bm{k}) = u_{\bm{k}} v_{\bm{k}} $ for various densities as shown for (a) zero detuning $ \omega = -E_0 $ (b) blue detuning $ \omega = 0 $.  (c) The singlet pair function peak momenta for detunings shown.  (d) The condensate 
interaction energy per
unit density for various chemical potentials.}
\end{figure}

Given that at high density there is inevitably a large number of photons, we may return to the original Hamiltonian (\ref{grandham}) to see the consequences. In the high density limit we may replace the photon operator by a classical c-number, which we consider to be very large $ a \approx \lambda \gg 1 $. Discarding all terms that do not contain this factor, we have
\begin{align}
\label{hidensityham}
H \approx \Omega \lambda \sum_{\bm{k}} \left[ e_{\bm{k}}^\dagger h_{-\bm{k}}^\dagger  +  h_{-\bm{k}} e_{\bm{k}} \right] + \lambda^2  \omega .
\end{align}
The solution of this Hamiltonian is the BCS wavefunction (\ref{bcswavefunction}) with $ u_{\bm{k}} = - v_{\bm{k}} = 1/\sqrt{2} $, agreeing with the numerical analysis that the average 
occupation number approaches $ 1/2 $ for all $ k $. The BCS gap in this limit corresponds to a single excitation of the Hamiltonian, which has an energy
\begin{equation}
E_{\bm{k}}^{\mbox{\tiny min}} \approx 2 \Omega \lambda .
\end{equation}
Thus with increasing density the BCS gap continues to increase in agreement with Fig. \ref{fig1}d. 

Now we ask to what extent the polariton system possesses BCS features, rather than merely a crossover between a polariton BEC state to a photon-dominated regime. The criterion given in Ref. \cite{keeling05} is 
based on a comparison of the energy scales of the BCS gap energy $ E_{\bm{k}}^{\mbox{\tiny min}} $ with the temperature required for condensation to occur $ k_B T_{\mbox{\tiny BEC}} = \pi \hbar^2 n/m_{\mbox{\tiny pol}}  $, where $ m_{\mbox{\tiny pol}} $ is the polariton mass. In this definition, the state can be described as ``BCS-like'' if the energy to disassociate a polariton is lower than the thermal excitation energy to prevent a BEC from occurring. In our units, this gives $ k_B T_{\mbox{\tiny BEC}}/E_0 \approx 22000 n a_B^2  $, 
where we used $ m_{\mbox{\tiny pol}} \approx 10^{-5} m_0 $ and $ a_B = 10$nm, and $ m_0 $ is the free electron mass. In terms of Fig. \ref{fig1}d, 
this criterion is always much higher than the gap energy $ E_{\bm{k}}^{\mbox{\tiny min}} $, for densities exceeding $n \approx 6.5 \times 10^7 $ $\mbox{cm}^{-2} $.  Experimentally, such densities have already been achieved, giving the result that all current polariton BECs are all in the  ``BCS-like'' regime, 
according to this criterion.  

There is however another sense that the polaritons can be classified as BCS-like.  
For an excitonic BCS state, the singlet pair function $ \Psi(\bm{k}) = u_{\bm{k}} v_{\bm{k}} $
is peaked near the vicinity of the Fermi momentum and has a width of the order of the inverse of the BCS coherence \cite{waldram96}. Figure \ref{fig2}a shows the singlet pair function for our polariton system, which is peaked at non-zero momentum for large densities.  The location of this peak has a non-zero momentum 
above a critical density (Fig. \ref{fig2}c). Such behavior is also seen in the excitonic BEC-BCS crossover. The difference here is that instead of the singlet 
pair function becoming sharper with increasing density, here the pair wavefunction becomes broader. Using blue-detuned (more excitonic) polaritons $ \Psi(\bm{k}) $ more resembles the excitonic BCS state (Fig. \ref{fig2}b). An alternative definition of a polariton BCS phase may be signaled by presence of peak of the singlet pair function at non-zero momentum. 

\begin{figure}
\scalebox{0.3}{\includegraphics[bb=0 0 1345 688]{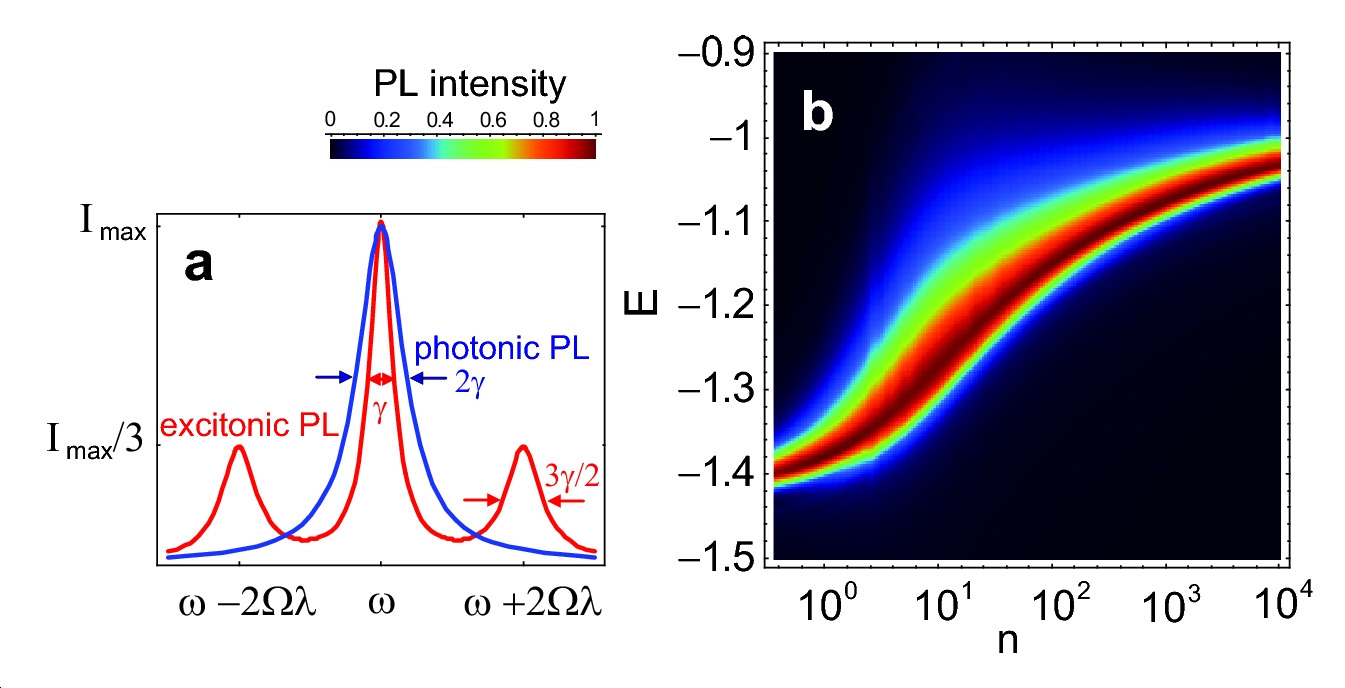}}
\caption{\label{fig3}
(a) The PL intensity of the BCS model in the high density limit for the photonic PL and 
excitonic PL. A reservoir coupling $ \gamma $ is assumed for both cases, giving the linewidths as shown. (b) The photonic PL for the BCS model at all densities. Zero detuning is assumed in all cases.}
\end{figure}

We now turn to the photoluminescence (PL) characteristics of the transition between low and high density.  Examining the high density limit first, using a similar approximation to (\ref{hidensityham}), we use the Hamiltonian
\begin{align}
\label{mollowham}
H =  \Omega \sum_{\bm{k}} \left[ \sigma^+_{\bm{k}} a + \sigma^-_{\bm{k}} a^\dagger \right] +  \omega a^\dagger a + \frac{\epsilon_{\mbox{\tiny g}}}{2} \sum_{\bm{k}} \sigma^z_{\bm{k}},
\end{align}
where $ \sigma^+_{\bm{k}} = e^\dagger_{\bm{k}} h^\dagger_{-\bm{k}} $, $ \epsilon_{\bm{k}} = \hbar^2 k^2/m $, 
$ \sigma^z_{\bm{k}} = e^\dagger_{\bm{k}} e_{\bm{k}} = h^\dagger_{-\bm{k}} h_{-\bm{k}} $, and we have explicitly included the semiconductor band gap energy $ \epsilon_{\mbox{\tiny g}} $ required to create an electron hole pair. The PL spectrum is calculated
using \cite{scully97}
\begin{equation}
\label{pldefinition}
I(\epsilon) = \frac{1}{\pi} \mbox{Re} \int_0^\infty \langle A^\dagger (\tau) A (0)  \rangle
e^{-i \epsilon \tau/\hbar} d \tau
\end{equation}
where $ A $ is the operator that couples the system to the external PL field. In the case of polaritons, the PL
is generally observed by leakage of the photon field through the microcavity, hence $ A = a $. It is also possible to observe the PL via a secondary means, from the coupling to the exciton field $ A = \sigma^-_{\bm{k}} $. 
This type of coupling is that measured for pure excitons and should also be present in principle for polaritons.
Experimentally the excitonic PL is emitted homogeneously in all directions, whereas the photonic PL is emitted perpendicularly to the sample surface. Evaluating the time correlation function (\ref{pldefinition}) for both types of couplings under a mean field approximation, we find a spectrum as shown in Fig. \ref{fig3}a. For the exciton coupling, the PL spectrum is 
identical to the familiar Mollow's triplet spectrum found in resonant fluorescence \cite{scully97}. For the photon coupling
only the central peak is present. The reason for this difference is illustrated by the single spin version of (\ref{mollowham}), which has eigenstates for high density $ |\pm,N \rangle = ( |\uparrow,N-1 \rangle \pm |\downarrow,N \rangle)/\sqrt{2} $, where $ N $ is the number of photons. For large $ N $, the 
photon operator does not cause transitions between the $ \pm $ eigenstates: $ a |\pm,N \rangle \approx \sqrt{N} |\pm,N-1 \rangle $. In contrast, the exciton coupling does cause a transition $ \sigma^- |\pm,N \rangle \approx (|+,N-1 \rangle - |-,N-1 \rangle )/\sqrt{2} $, giving the side peaks.

In the low density limit, either type of coupling gives the same PL spectrum, giving the familiar emission centered at the lower polariton energy.  For the photon coupling case, in both the low and high density limits, the action of applying the operator $ A = a $ does not cause transitions to excited states. Assuming this is true for all intermediate densities, we use the results of our ground state BCS wavefunction to obtain the PL emission parameters.  From the results of the self-consistent equations (\ref{bcs1})-(\ref{bcsequations}), we
may write down an effective Hamiltonian for the BCS theory $ H_{\mbox{\tiny eff}} = \mu b^\dagger b + \frac{1}{2} V b^\dagger b^\dagger b b  $, where $ b $ is a bosonic operator for the effective theory, $ \mu = \frac{dE}{dn} $ and $ V = \frac{1}{A} \frac{d^2E}{dn^2} $, where $ A $ is the sample area and $ E $ is an energy density.  The peak energy of the PL is the energy of adding a single particle to the system, which is by definition equal to the chemical potential $\mu $. To determine the 
linewidth of the spectrum, we use the method presented in Ref. \cite{porras03} to incorporate the 
effect of self-interaction on the PL. The mean field self-interaction energy of the condensate $ n V$ is shown in Fig. \ref{fig2}d, which shows that interaction reaches a maximum 
at intermediate density. We attribute this to the fact that at high density the particles are more likely to be 
present in a photonic state, which have no interactions with other particles. Inputting the interaction energy into the theory of Ref.  \cite{porras03} gives the PL spectrum Fig. \ref{fig3}b. The PL spectrum gradually shifts from the lower polariton energy to the cavity photon energy, with a asymmetric linewidth with an exponential tail towards high energy. The linewidth decreases again in the high density limit, due to the decreased interactions. 

In the excitonic PL, the side peaks of the Mollow's triplet should disappear at intermediate densities, when the saturation effect of the excitons becomes negligible. The central peak of the excitonic PL should exhibit a similar behavior to that shown in Fig. \ref{fig3}b. We note that only the zero center of mass momentum PL
is considered in our analysis and we leave calculations of dispersion characteristics as future work. 

We have analyzed the crossover between low and high density limits of exciton-polariton condensates using a BCS wavefunction approach. Contrary to the exciton case, the electron hole pairs have a reduced separation in the high density limit due to the dominant cavity photon field.
Intuitively we picture this state as a strong cavity photon field continuously creating and destroying electron hole pairs at localized positions in the quantum well, resulting in a half occupancy of $ v_{\bm{k}}^2 $. 
In the intermediate density regime, the system has BCS-like properties in the sense that the pair breaking energy is less than the energy required for destroying the condensate, and a peak in the singlet pair function develops.  
The photonic PL shifts from the lower polariton energy towards the cavity photon energy with a broadening of the linewidth due to the increased interactions. In the high density limit the excitonic PL should exhibit a Mollow's triplet type structure, originating from the saturation of the electron-hole occupancy. Inclusion of non-equilibrium effects resulting from a coupling to an external bath is left as future work. We expect that inclusion of this effect will reduce the energy gap due to pair-breaking dephasing processes \cite{szymanska02}. Signatures of the high density regime should be observable in current experimental systems. 

This work is supported by the Special Coordination Funds for Promoting Science and Technology, Navy/SPAWAR Grant N66001-09-1-2024, MEXT, and NICT.


\end{document}